\documentclass[preprint,12pt,authoryear]{elsarticle}

\usepackage{amssymb}
\usepackage{amsmath}
\usepackage{amsthm}
\usepackage{graphicx}
\usepackage{subfig} % for side by side figures
\usepackage{bbm} % for the bold fancy one symbol
\usepackage{enumerate}
\usepackage[ruled]{algorithm2e}
\usepackage{url} % not crucial - just used below for the URL 
\usepackage{float}
\usepackage{xcolor}
\usepackage{comment}
\usepackage{multirow}

\journal{Computational Statistics \& Data Analysis}
\input{format.tex}

\begin{document}

\begin{frontmatter}



\author[GMU]{Nicholas Rios} %% Author name
\author[GMU]{Seiyon Lee\corref{slee}}

\affiliation[GMU]{organization={George Mason University},%Department and Organization
            addressline={4400 University Drive}, 
            city={Fairfax},
            postcode={22030}, 
            state={VA},
            country={United States}}

\cortext[slee]{Corresponding author. \textit{Postal address:} 4400 University Drive, Fairfax, VA 22030, USA. 
\textit{Email:} slee287@gmu.edu. 
\textit{Tel:} +1-703-993-3505. 
\textit{Fax:} +1-703-993-1700.}

\title{REX-SUB: A Scalable Subsampling Strategy for Modeling Large Spatial Datasets}

\bigskip
\begin{abstract}
Recent advances in data collection technologies have led to the emergence of massive spatial datasets, with measurements obtained at millions of spatial locations. Geostatistical models typically employ Gaussian processes (GPs) to capture spatial dependence, but standard GP fitting becomes prohibitive at such scales. A promising solution is optimal subsampling, where a subset of locations is selected that optimizes a criterion. In this study, we propose a randomized exchange algorithm for subsampling (REX-SUB) which efficiently selects small subsamples that minimize prediction errors in the fitted spatial GP models. To further improve computational efficiency, we embed a scalable Vecchia approximation to the GP's joint likelihood, which takes advantage of sparsity in the precision matrix to enable fast inference on the selected subsamples. Through a simulation study and an application to a remotely sensed precipitable water dataset, we show that REX-SUB yields lower mean squared prediction errors and interval scores compared to competing subsampling strategies.
\end{abstract}

%% Keywords

\end{frontmatter}

\newpage

\section{Introduction}
\label{sec:intro}
 
Recent developments in data collection technologies, such as remote sensing, wearable devices, and citizen science initiatives have led to the emergence of massive spatial datasets, often comprising millions of locations. These datasets play an integral role across numerous disciplines and applications, including examples such as remotely sensed aerosol optical depth \citep{wei2019modis}, cloud cover \citep{sengupta2013hierarchical}, total precipitable water \citep{zilber2021vecchia}, counts of bird species in ecological surveys \citep{guan2018computationally}, and ice thickness measurements in glaciological studies \citep{fretwell2013bedmap2}. In addition to their massive scale (millions of locations), these spatial datasets often exhibit strong spatial dependence, which must be properly modeled to improve predictions at unobserved locations and to quantify associated uncertainties.

Geostatistical models provide a flexible framework for spatial data analysis by modeling the underlying spatial process as a latent Gaussian process (GP) with spatially varying mean and covariance functions \citep{williams2006gaussian, wahba1990spline}. This approach captures spatial dependence through the structure of the Gaussian random field \citep{cressie2011statistics, banerjee2003hierarchical}. For a finite set of locations, a GP imposes a multivariate normal distribution with a positive definite covariance matrix that can reflect nonstationarity, anisotropy, or heteroskedasticity. However, likelihood evaluations can be costly due to the inherent matrix inversions and determinant calculations. As a result, standard GP models become computationally infeasible for large datasets (e.g., $N > 10^4$), due to the $\mathcal{O}(N^3)$ cost of dense matrix operations like Cholesky decomposition.

Scalable methods have been developed to address these computational challenges. Low-rank approaches represent the latent spatial random process as an expansion of spatial basis functions, such as process convolutions \citep{higdon1998process}, fixed-rank kriging \citep{cressie2008frk}, and predictive processes \citep{Banerjee2008Gaussian}. However, the resulting low-rank approximations may oversmooth spatial predictions \citep{stein2014limitations}. Alternatively, sparse matrix methods approximate the covariance or precision structure using sparse representations, which are amenable to large matrix operations. Covariance tapering \citep{furrer2006covariance} applies a tapering function to induce sparsity in the covariance matrix while maintaining positive definiteness. 
The stochastic partial differential equations (SPDE) approach \citep{lindgren2011explicit} characterizes the latent GP as a solution to stochastic partial differential equations, which yields sparse precision matrices. The Vecchia approximations \citep{Vecc:1988, katzfuss2020vecchia} assume that the joint conditional distribution can be approximated by conditioning on a smaller conditioning set of size $m\ll n$. This leads to sparse precision matrices or sparse Cholesky factors \citep{datta2016hierarchical, katzfuss2020vecchia, katzfuss2021general}, enabling fast likelihood evaluations. Although these approaches scale well, they still require modeling the full dataset, which remains computationally demanding for massive spatial datasets. 

Spatial subsampling methods have been proposed to reduce the computational burden of modeling large spatial datasets. These approaches typically divide the domain into blocks using equal-area partitioning \citep{sang2011covariance}, hierarchical clustering \citep{heaton2019case,lee2023scalable}, or tree-based methods \citep{konomi2014adaptive}, and often assume block-wise independence \citep{saha2024incorporating,bradley2015comparing}. A notable exception is the `spatial data subset model' \citep{saha2024incorporating,bradley2021approach}, which embeds simple random subsampling within a Bayesian hierarchical framework. Here, the model parameters are estimated using randomly selected subsets rather than the larger full dataset. However, it does not incorporate optimality criteria for selecting the most informative subsets, nor does it explore the optimal subsample size needed for inference.

The computational challenges brought about by massive datasets are not unique to spatial statistics, and are seen in a wide variety of disciplines in science and industry. This motivated the development of optimal subsampling, which is the practice of selecting a small subset of data points (referred to as the subsample) from a large dataset that contains the most useful information for fitting a particular model. This way, the model of interest can be efficiently fitted to the small subsample without loss of critical information. A notable method is Information Based Optimal Subdata Selection (IBOSS), which chooses subdata that maximize the determinant of the Fisher information matrix; this is inspired by the $D-$optimality criterion in experimental design \citep{cook1980comparison, jones2021optimal}. IBOSS was first implemented for linear regression models \citep{wang2019information}, and then it was extended to logistic regression \citep{cheng2020information}, LASSO frameworks \citep{singh2023subdata}, and cluster-wise linear regression \citep{liu2023information}. \cite{shi2021model} also studied an approach for subsampling that is robust to model misspecification by finding subsamples based on the maximin distance criterion typically employed in space-filling designs \citep{johnson1990minimax}. However, the body of literature on optimal subsampling does not emphasize criteria that yields high-quality predictions over a region of interest. It also does not cover spatial statistical models. 

This paper presents REX-SUB, a randomized exchange algorithm that enables scalable and efficient subsample selection from massive spatial datasets. Two optimality criteria will be proposed for subsample selection; the first minimizes prediction errors, while the second minimizes interval scores. The proposed algorithm will allow for a flexible class of spatial Gaussian Process (GP) models to be fitted on a carefully curated subsample. We demonstrate and validate REX-SUB through a comprehensive simulation study, where the proposed methodology and competing methods are validated on various spatial dependence settings. Rather than a novel GP approximation, REX-SUB is a practical subsampling heuristic designed for computational efficiency while preserving strong inferential and predictive performance in large spatial datasets.

The rest of this paper is organized as follows. Section \ref{sec:prelims} will review existing methods for fitting Geostatistical models and performing subsampling for GPs. Section \ref{sec:methods} will introduce the proposed criteria for optimal subsamples and the REX-SUB method. Section \ref{sec:simuresults} will show simulation results that demonstrate the efficacy of REX-SUB. In Section \ref{sec:example}, REX-SUB will be applied to a remotely-sensed precipitable water dataset collected NASA's Terra satellite. Section \ref{sec:conc} concludes the paper.

\section{Preliminaries}
\label{sec:prelims}

\subsection{Geostatistical Models}
\label{subsec:spatial_models}

In geostatistics, Gaussian process (GP) models are widely used to model spatially correlated processes. GPs provide a flexible and probabilistic approach for modeling spatial dependence, allowing for both interpolation at unobserved locations (kriging) and uncertainty quantification, particularly for predictions. The spatial dependence structure is defined through a covariance function that defines how correlation between observations decays with the distance between locations; this makes GPs particularly well-suited for modeling spatial data across many fields. 

Let ${Z(\bs)}$ denote a spatial random process indexed by locations $\bs\in \mathcal{D}\subset \mbR^d$ where $d$ is the spatial dimension, which is typically 2 or 3 dimensions. In geostatistical models, it is assumed that:
\begin{equation}
\label{eq:GPModel}
Z(\bs)\sim \mathcal{GP}(\mu(\bs), K(\bs, \bs^{\prime}; \bTheta)),
\end{equation}
where $\mu(\bs)= \mbE[Z(\bs)]$ is the spatially varying mean function, $K(\bs, \bs^{\prime}; \bTheta)$ is the covariance function with covariance parameter vector $\bTheta$, and $\bs$ and $\bs^{\prime}$ denote two locations. A commonly used specification for $\bTheta$ stems from the Mat\'ern class of covariance functions with parameters $\bTheta=(\nu,\phi,\sigma^2,\tau^2)^{\prime}$ with smoothness $\nu$, range $\phi$, partial sill $\sigma^2$, and nugget variance $\tau^2$: 

\begin{equation}
\label{eq:matern}
K(\bs, \bs' ; \boldsymbol\Theta) = \sigma^2 \cdot \frac{2^{1 - \nu}}{\Gamma(\nu)} 
\left( \frac{\sqrt{2\nu} \, \|\bs - \bs'\|}{\phi} \right)^\nu 
K_\nu\left( \frac{\sqrt{2\nu} \, \|\bs - \bs'\|}{\phi} \right) +\tau^2\mathbb{I}(\bs= \bs').
\end{equation}

For a finite set of locations, consider spatially indexed observations $\{Z(\bf{s}):\bf{s} \in\mathcal D\}$ observed at locations $\bf{s}\in\mathcal D\subset\mathbb R^2$. The observations are collected at a finite set of $N$ locations such that $\bz=\{Z(\bf{s}_1),\ldots, Z(\bf{s}_N)\}$, and $\bz \sim \mathcal{N}(\bmu, \bK_\Theta)$ for a spatially varying mean function $\bmu= \{\mu(\bf{s_1}),\ldots,\mu(\bf{s_N})\}$ and covariance matrix $\bK_\Theta$. 

Kriging is used for optimal predictions at unobserved locations by leveraging the spatial correlation structure embedded within Gaussian process models as well as uncertainty quantification via the predictive covariance matrices. Under the assumed Gaussian process model, the kriging estimator corresponds to the best linear unbiased predictor \citep{cressie2011statistics} often used to predict at unobserved values under a linear mixed modeling framework. Given the inferred covariance parameters $\bTheta$ and observation $\bz$, the predictive distribution of $Z(\bs_0)=z_0$ for any $\bs_0  \in \mathcal{D}$ is:
\begin{equation}
z_0 \mid \bz, \bTheta \sim \mathcal{N}\left( \mu_0 + \bK_{\Theta[0,1]} \bK_{\Theta[1,1]}^{-1}(\bZ - \bmu), \; \bK_{\Theta[0,0]} - \bK_{\Theta[0,1]} \bK_{\Theta[1,1]}^{-1} \bK_{\Theta[1,0]}\right),
\end{equation}
with mean functions $\mu_0$ and $\bmu$ at the unobserved and observed locations, respectively, and covariance matrix  $\bK_{\Theta}=\begin{bmatrix}
\bK_{\Theta[0,0]}& \bK_{\Theta[0,1]}\\
\bK_{\Theta[1,0]}& \bK_{\Theta[1,1]}\\
\end{bmatrix}$. 

Though GPs serve as versatile building blocks for geostatistical models, they are subject to key computational limitations that impact their applicability to real-world problems. The main challenge is rooted in their computational complexity  $\mathcal{O}(n^3)$ and memory requirement $\mathcal{O}(n^2)$ stemming from the need to invert and store the dense covariance matrix \citep{williams2006gaussian}. The computational costs render standard GP models computationally prohibitive for even moderately large spatial datasets (thousands of locations). In addition, spatial interpolation (kriging) often requires $\mathcal{O}(nm)$ computational operations, which can be difficult to perform  when the number of observed $(n)$ or prediction locations $(m)$ are large. Moreover, the covariance function parameters $\bTheta$ must be carefully tuned based on the observed data. When the sample size $n$ is large, parameter tuning and optimization can become computationally prohibitive, often resulting in poor convergence and model misspecification, which in turn can lead to inaccurate and imprecise predictions.

\subsection{Vecchia Approximations for Scalable Spatial Inference}
\label{subsec:vecchia}
Vecchia approximations  \citep{vecchia1988estimation,stein2004approximating} are a scalable approach for approximating the likelihood of high-dimensional Gaussian processes. 
Computational efficiency is achieved by factorizing the joint density function into a sequence of low-dimensional conditional densities. This factorization yields a sparse representation of the Cholesky factor of the precision matrix, considerably reducing the computational costs of large-scale matrix operations.

For the spatial process $Z(\bs)$ in (\ref{eq:GPModel}), the standard Vecchia approximation expresses the joint density function as:
\begin{equation}
\label{eq:Vecchia}
\tilde{f}(\mathbf{z}) = f(z_1) \prod_{i=2}^n f(z_i \mid z_{\mathcal{N}(i)}),
\end{equation}
where $ \mathcal{N}(i) \subset \{1, \dots, i-1\} $ is a small conditioning set, often defined as the $m$ nearest neighbors of location $\bs_i$. This approximation assumes that the full conditionals $ f(z_i \mid z_{1:(i-1)})$ can be adequately approximated by conditioning only on $z_{N(\bs_i)}$. The neighborhood sets $N(\bs)=\{N(s_i):i=1,2,...,n\}$ are constructed from a directed acyclic graph $G=\{\mathcal{S},N_\mathcal{S}\}$ where $\mathcal{S}=\{\bs_1,\bs_2,...,\bs_n\}$ denotes the ordered locations. For each location $\bs_i$, the conditioning set $N(\bs_i) = \{\bs_{j_1}, \dots, \bs_{j_m}\}$ is chosen such that $ j_k < i$ for all $k$, preserving the directed nature of the graph.

The density in (\ref{eq:Vecchia}) defines a valid likelihood for any ordering of locations and any choice of conditioning sets. The ordering within $\bz$ is typically derived from the ordering of spatial locations $\bs_i$, using strategies such as coordinate-based (e.g., left-to-right or lower-left to upper-right) or maximum-minimum-distance (maxmin) ordering \citep{guinness2018permutation}. In maxmin ordering, the next location is chosen to maximize its minimum distance to all previously selected points. This strategy is often preferred in high-dimensional settings, as it ensures good spatial coverage of conditioning sets and leads to more accurate approximations \citep{zilber2021vecchia}.

Under the standard Vecchia model in (\ref{eq:Vecchia}), the resulting distribution is multivariate normal, $\hat{f}(\bu) = \mathcal{N}(\bzero, \bQ^{-1})$, with precision matrix $\bQ = \bU \bU^\top$, where $\bU$ is a sparse upper-triangular Cholesky factor. As demonstrated in the \texttt{GpGp} package, $\bU$ denotes the Cholesky factor of the precision matrix $\bQ$ with the reverse row-column ordering. The entries of $\bU$ are directly computed from the covariance $\bK_\Theta$. Please see the appendix of \citep{zilber2021vecchia} for details on the construction of $\bU$. 

Inference in \texttt{GpGp} proceeds by maximizing the Vecchia-approximated Gaussian log-likelihood using a single pass algorithm for computing the gradient and Fisher information \citep{guinness2021gaussian}, for which the Vecchia factorization enables efficient computation of the score function and Fisher information. Prediction is then carried out by extending the Vecchia factorization to include the prediction sites, allowing fast approximate kriging in which conditional means and variances are obtained from small nearest-neighbor systems rather than the full covariance matrix.

\subsection{Subsampling Approaches}
\label{subsec:subsampling}
\begin{figure}
    \centering
    \includegraphics[width=0.9\linewidth]{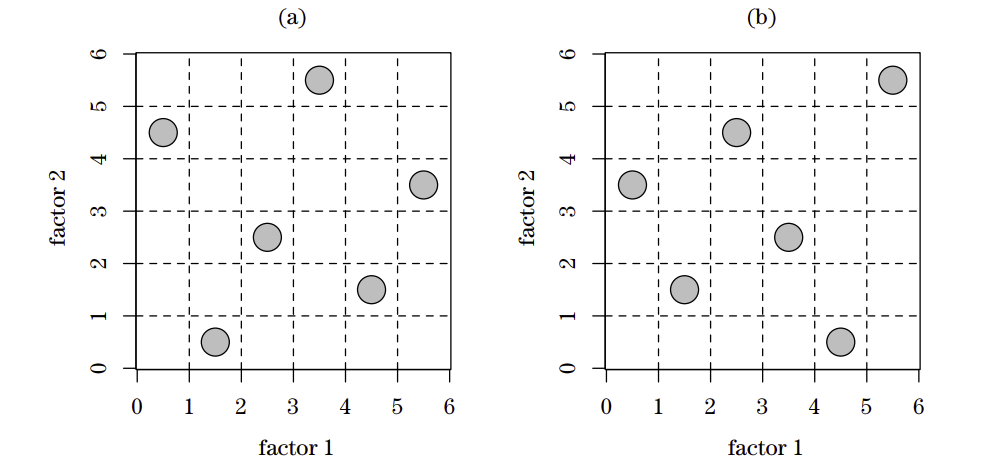}
    \caption{Two LHDs with $n = 6$ points in $d = 2$ dimensions. Figure from \cite{zhao2018efficient}.}
    \label{fig:LHS}
\end{figure}

Several approaches for subsampling have been proposed for Gaussian Process models. One of the main motivations for fitting a GP to a subsample (instead of a very large dataset) is to avoid costly matrix operations on the covariance matrix. The simplest approach is random subsampling, which is just selecting $n$ rows from a dataset of $N$ rows without replacement. The properties of GPs under random subsampling were extensively studied by \citep{hayashi2020random}, who derived theoretical bounds on their predictive mean and variance errors using graphons. Other existing approaches for GP subsampling are inspired by Latin Hypercube Designs (LHDs), which are a class of space-filling experimental designs introduced by \cite{mckaya1979comparison}. Suppose that each dimension of the experimental region (a subset of $\mathbb{R}^d$) is partitioned into $n$ equally sized blocks. When an LHD is projected onto any dimension, there is exactly one observation in each of the blocks. To construct an $n-$run LHD, one can permute the values $(1,2,\dots,n)$ in each of the $d$ dimensions, and then use these permuted values to select blocks in each dimension. A random point can then be sampled from each block by randomly generating a continuous Uniform random variable. The set of $n$ randomly sampled points is called a Latin Hypercube Sample (LHS). There are $(n!)^{d-1}$ possible LHD sampling schemes. Two examples of LHDs for $n = 6, d = 2$ are shown in Figure \ref{fig:LHS}. LHDs can also be generated from specific permutations that attempt to maximize the minimum distance between points; this is done by the \texttt{maximinLHS} function in the R package \texttt{lhs} \citep{lhspackage}. 

Naturally, LHDs can be used to subsample $n$ points from $N$ points in $d-$dimensional space by partitioning the space into $n \times n$ regions, and then using the LHD to select $n$ regions. Once the $n$ regions are selected, a random point (from the set of $N$ points) that belongs to the region is randomly sampled. In the event that no points from the full dataset fall in the region, another LHD can be generated. A similar approach (used in Section \ref{sec:simuresults}) is to generate an LHS of size $n$; then, for each point in the LHS, randomly sample one of the $N$ points in the full dataset that is within a small distance of this point. LHDs have also been used by \cite{zhao2018efficient} and \cite{he2022gaussian} to generate repeated subsamples for bootstrap aggregation of GPs. The key differences here are that \cite{zhao2018efficient} and \cite{he2022gaussian} used all observations in each of the $n$ randomly selected regions as a subsamples (as opposed to a single point), and repeated subsamples were taken to find the predicted mean and variances. 

An alternative to LHD subsampling is to select a subsample that minimizes a prediction-based criterion. A popular example is the Integrated Mean Squared Prediction Error (IMSPE), which is defined as \citep{santner2018design}
\begin{align}
    \text{IMSPE} = \int_{\textbf{s} \in D} \frac{E[(\hat{Z}(\textbf{s})-Z(\textbf{s}))^2]}{\sigma^2} \, d\textbf{s},
\end{align} where $D \subset \mathbb{R}^2$ is the space of locations. Lower values of the IMSPE are desirable, since they correspond to lower expected prediction error over the space of interest. If $D$ is rectangular, then the IMSPE has a closed form. Computation of the IMSPE requires knowledge of the hyperparameters of the covariance function used to make predictions over the location space, which are generally unknown prior to fitting the model. To overcome this difficulty, \cite{binois2021hetgp} proposed a sequential design approach, which we implement for comparison to REX-SUB later in Sections 4 and 5. In this approach, an initial subsample of size $n' < n$ is found using an LHD subsample, and an initial GP is fit to this subsample. Then, the location $\textbf{s}_{n'+1}$ in the subsample that minimizes the IMSPE criterion is selected and added to the current subsample. Next, the initial GP is updated using the data point $(\textbf{s}_{n'+1},Z(\textbf{s}_{n'+1}))$. This iterative process is repeated until the target subsample size $n$ is reached. As discussed in \cite{binois2021hetgp}, the location added at each iteration can be either a new location (not in the current subsample, but in the full sample), or a replicate of a location in the current subsample.

\section{Proposed Methods}
\label{sec:methods} 
Suppose that we have data $\{(\textbf{s}_i, z_i), i = 1,\dots,N\}$ where $\textbf{s}_i \in \mathbb{R}^2$ are coordinates of spatial locations where samples are taken and $z_i \in \mathbb{R}$ are univariate responses, which correspond to measurements taken at each location. Consider the Spatial GP model \begin{align}
    \label{eqn:SGPmodel}
    \textbf{Z} \sim \mathcal{N}(\boldsymbol\mu, \bK({\boldsymbol\Theta})),
\end{align} where $\boldsymbol\mu = [\mu(\textbf{s}_1), \dots, \mu(\textbf{s}_N)]^T$. It is assumed that $\boldsymbol\mu = \textbf{0}$ because the spatial responses can be mean-centered prior to fitting any model. The covariance matrix $\bK({\boldsymbol\Theta})$ has its entries determined by a covariance function $K(\cdot, \cdot; \boldsymbol\Theta)$ with hyperparameters $\boldsymbol\Theta$.

If the number of observations $N$ is quite large, then Model (\ref{eqn:SGPmodel}) is computationally expensive to fit. This motivates taking a subsample of $n$ locations from the set of $N$ possible locations. Let $\mathcal{I}$ be a set of indices that corresponds to a subsample. One can then fit Model (\ref{eqn:SGPmodel}) to the subsample. Let $\hat{\boldsymbol\Theta}_{\mathcal{I}}$ be the parameter estimates obtained from fitting Model (\ref{eqn:SGPmodel}) to $\{ (\textbf{s}_i, Z(\textbf{s}_i)) \mid i \in \mathcal{I}\}$. Then, predictions can be made for any spatial location $\textbf{s} = (s_1,s_2)$. Let $\textbf{Z}_{\mathcal{I}} = [Z(\textbf{s}_i)]_{i \in \mathcal{I}} \in \mathbb{R}^n$, $\textbf{k}(\textbf{s};\hat{\boldsymbol\Theta}_{\mathcal{I}}) = [K(\textbf{s},\textbf{s}_i; \hat{\boldsymbol\Theta}_{\mathcal{I}})]_{i \in \mathcal{I}} \in \mathbb{R}^n$, and let $\bK(\hat{\boldsymbol\Theta}_{\mathcal{I}}) = [K(\textbf{s}_i, \textbf{s}_j; \hat{\boldsymbol\Theta}_{\mathcal{I}})]_{i \in \mathcal{I}, j \in \mathcal{I}} \in \mathbb{R}^{n \times n}$. Then, given the fitted GP from subsample $\mathcal{I}$, the predicted value of the response at any location $\textbf{s}$ is \begin{align}
    \label{subsamplepredz}
    \hat{Z}_{\mathcal{I}}(\textbf{s}) = \textbf{k}(\textbf{s}; \hat{\boldsymbol\Theta}_{\mathcal{I}})^T[\bK(\hat{\boldsymbol\Theta}_{\mathcal{I}})]^{-1}\textbf{Z}_{\mathcal{I}}.
\end{align} Conditional on the subsample, the estimated variance of $\hat{Z}_{\mathcal{I}}(\textbf{s})$ is \begin{align}
    \label{eqn:varpredz}
    \widehat{\text{Var}}[\hat{Z}_{\mathcal{I}}(\textbf{s})] = K(\textbf{s},\textbf{s}; \hat{\boldsymbol\Theta}_{\mathcal{I}}) - \textbf{k}(\textbf{s}; \hat{\boldsymbol\Theta}_{\mathcal{I}})^T[\bK(\hat{\boldsymbol\Theta}_{\mathcal{I}})]^{-1}\textbf{k}(\textbf{s}; \hat{\boldsymbol\Theta}_{\mathcal{I}})^T,
\end{align} which can be used to construct prediction intervals for $Z(\textbf{s})$. While one could always randomly select the subsample, this might not always give predictions that are accurate or have low variance. We propose several optimality criteria for evaluating the ``quality'' of a particular subsample. Let $\phi(\mathcal{I} \mid \textbf{S}, \textbf{Z})$ be an optimality criterion that takes the indices of a subsample $\mathcal{I}$ and returns a real number. All considered optimality criteria are written so that smaller values of $\phi$ correspond to better subsamples. For example, one can be interested in choosing subsamples of the spatial locations that have low prediction error. In this case, $\phi$ can be defined as \begin{align}
    \label{eqn:phiMSPE}
    \phi_{\text{MSPE}}(\mathcal{I} \mid \textbf{S}, \textbf{Z}) \triangleq \frac{1}{|\mathcal{I}_{test}|}\sum_{i \in \mathcal{I}_{test}}(\hat{Z}_{\mathcal{I}}(\textbf{s}_i) - Z(\textbf{s}_i))^2,
\end{align} where $\mathcal{I}_{test} \subset \{1,\dots,N\}$ is a subset of indices for spatial locations used to test prediction accuracy. The spatial locations in $\mathcal{I}_{test}$ are randomly selected from all available locations before any optimal subsample selection is performed. Once chosen, the set $\mathcal{I}_{test}$ is separated from the set of all available locations. In other words, the proposed methods only fit GPs to subsamples that are subsets of $\{1,\dots,N\} \setminus \mathcal{I}_{test}$. This is done to ensure that there is no overlap between the sample that a GP is trained on and the held-out sample used to evaluate criteria like $\phi_{MSPE}$. It is also of interest to use criteria that also consider the variance of the predictions $\hat{Z}_{\mathcal{I}}(\textbf{s})$. One such criterion is the Interval Score (IS) discussed in \cite{gneiting2007strictly}, which evaluates the quality of an interval estimator. The Interval Score is a quantitative measure that reflects the uncertainty of model predictions. For a given prediction interval, the Interval Score increases if the interval becomes wider and also increases if the interval does not contain the true predicted response. This measure has been used in many applications to evaluate the quality of prediction intervals in Gaussian Processes; for examples, see \cite{wood2022scalable}, \cite{yerramilli2023fully}, and \cite{chen2025geodgp}. Specifically, suppose that $(L_{\mathcal{I}}(\textbf{s}), U_{\mathcal{I}}(\textbf{s}))$ is a $100(1-\alpha)\%$ prediction interval for $Z(\textbf{s})$. In this context, the IS criterion can be written as \begin{equation}
    \label{eqn:phiIS}
    \begin{aligned}
        \phi_{IS}(\mathcal{I} \mid \textbf{S}, \textbf{Z}) &= \frac{1}{|\mathcal{I}_{test}|}\sum_{i \in \mathcal{I}_{test}} \Bigg( U_{\mathcal{I}}(\textbf{s}_i)-L_{\mathcal{I}}(\textbf{s}_i) \\
        &+ \frac{2}{\alpha}\Big(L_{\mathcal{I}}(\textbf{s}_i) - Z(\textbf{s}_i) \Big)I\Big(Z(\textbf{s}_i) < L_{\mathcal{I}}(\textbf{s}_i)\Big) \\
    &+ \frac{2}{\alpha}\Big(Z(\textbf{s}_i) - U_{\mathcal{I}}(\textbf{s}_i)\Big)I\Big(Z(\textbf{s}_i) > U_{\mathcal{I}}(\textbf{s}_i)\Big) \Bigg).
    \end{aligned}
\end{equation} If the prediction interval $(L_{\mathcal{I}}(\textbf{s}_i), U_{\mathcal{I}}(\textbf{s}_i))$ contains the true response $Z(\textbf{s}_i)$, then the interval score for test location $\textbf{s}_i$ is just the length of the corresponding prediction interval. Otherwise, the interval score for location $\textbf{s}_i$ is penalized if the prediction interval does not contain the observed response; the magnitude of this penalty is proportional to the distance between the nearest endpoint of the prediction interval to the observed response. The $\phi_{IS}$ criterion in (\ref{eqn:phiIS}) is just the average of the interval scores across all test locations. Therefore, lower values of $\phi_{IS}$ are more desirable, as they signify that the prediction intervals (on average) have lower uncertainty and are more likely to capture the predicted response.

It is difficult to deterministically select subsamples that will optimize the $\phi_{MSPE}$ or $\phi_{IS}$ criteria. The optimality criteria depend on parameter estimates $\hat{\boldsymbol\Theta}_{\mathcal{I}}$, which can change depending on the subsample. Instead, a randomized version of Federov's exchange algorithm  \citep{fedorov2013theory} is considered. Federov's algorithm iteratively exchanges points of an experimental design with a list of candidate points, and makes exchanges that improve an optimality criterion. In this case, the points correspond to locations in the current subsample, and the candidate points are randomly selected from the full list of available locations at each iteration. The proposed method for spatial subsampling is summarized in Algorithm \ref{Alg:optsubsample}.

%\spacingset{1}
\begin{algorithm}[H]
\label{Alg:optsubsample}
  \SetAlgoLined
  \textbf{Inputs}: $\textbf{S} = [\textbf{s}_1, \dots, \textbf{s}_N]^T$, $\textbf{Z} = [Z(\textbf{s}_1), \dots,Z(\textbf{s}_N)]$, subsample size $n$, number of random candidates $n_{cand}$, number of repetitions $n_{repeat}$, optimality criterion $\phi$.\\
  1. Randomly select $n$ subsample indices $\mathcal{I}^{best}$ from $\{1,2,\dots,N\}$.  \\  
  2. Use the Vecchia approximation to fit a GP to $\{(\textbf{s}_i,Z(\textbf{s}_i)) \mid  i \in \mathcal{I}^{best}\}$, and use it to find $\phi^{best} = \phi(\mathcal{I}^{best}, \textbf{S}, \textbf{Z}).$ \\ 
  \For{$r = 1,\dots,n_{repeat}$}{
          \For{$j = 1,\dots,n$}{   

            \For{$t = 1,\dots,n_{cand}$}{
                      3. Copy $\mathcal{I}^{(t)} = \mathcal{I}^{best}.$ \\
                            4. Randomly sample $j'$ from $\{1,\dots,N\}\setminus \mathcal{I}^{best}$. Change the $j^{th}$ element of $\mathcal{I}^{(t)}$ to $j'$. \\
                            5. Use the Vecchia approximation to fit a GP to $\{(\textbf{s}_i,Z(\textbf{s}_i)) \mid  i \in \mathcal{I}^{proposed}\}$, and use it to find $\phi^{(t)} = \phi(\mathcal{I}^{(t)}, \textbf{S}, \textbf{Z}).$ Store $\phi^{(t)}$. \\
                            
            }
            6. Set $t^*= \arg\min_{t = 1,\dots,n_{cand}} \phi^{(t)}$ and $\mathcal{I}^* = \mathcal{I}^{(t^*)}$. \\
            \If{$\phi^{(t^*)} < \phi^{best}$}{
                                7. $\mathcal{I}^{best} = \mathcal{I}^{(t^*)}$, $\phi^{best} = \phi^{(t^*)}$. 
                            }

          }  
  }

 \Return{$\mathcal{I}^{best}$.}
 \caption{Randomized Federov Exchange Algorithm for Spatial Subsampling (REX-SUB)}
\end{algorithm}

As inputs, Algorithm \ref{Alg:optsubsample} takes the matrix of $N$ spatial locations $\textbf{S}$, their associated responses $\textbf{Z}$, the desired subsample size $n < N$, a number of random candidates $n_{cand}$, a number of overall repetitions $n_{repeat}$, and an optimality criterion $\phi$ to evaluate the quality of the spatial subsample. The goal of the algorithm is to identify subsamples with low values of the optimality criterion. In Step 1, a random subsample of row indices is taken and stored as $\mathcal{I}^{best}$, i.e., the best set of indices discovered so far. In Step 2, Model (\ref{eqn:SGPmodel}) is fitted to the subsample corresponding to $\mathcal{I}^{best}$ and this is used to evaluate $\phi$. Algorithm \ref{Alg:optsubsample} then considers each location sequentially. In Steps 3 to 5, a list of $n_{cand}$ locations that are not currently in the subsample are randomly selected (with equal probability). The $j^{th}$ location of the subsample is then exchanged with each candidate location, and the new $\phi-$optimality is stored. In Step 6, the exchange that results in the best $\phi-$optimality is chosen. In Step 7, if this exchange lowers the optimality criterion, the exchange is made. Steps 3 to 7 are repeated a total of $n_{repeat}$ times. The algorithm returns the set of row indices that correspond to the spatial locations that led to the lowest optimality criterion found. 

Algorithm \ref{Alg:optsubsample} is not guaranteed to find a subsample that globally minimizes $\phi$. This is because Algorithm \ref{Alg:optsubsample} is ``greedy'' in the sense that it only makes nearby exchanges if they strictly reduce $\phi$. However, when $\phi$ is the  MSPE criterion in (\ref{eqn:phiMSPE}), simulations in Section \ref{sec:simuresults} and real data examples in Section \ref{sec:example} show that this method results in samples with lower MSPE than both random sampling and subsamples found based on Latin Hypercube Sampling (LHS).

REX-SUB requires two main tuning parameters. The first is the number of random candidate locations $n_{cand}$, and the second is the number of repetitions $n_{repeat}$. Increasing either of these tuning parameters will generally result in a more optimal subsample, at the cost of increased computation time. Increasing $n_{cand}$ allows REX-SUB to explore more locations at each iteration, which can lead to better subsamples. Fortunately, Steps 3 to 5 of Algorithm 1 can be parallelized. This is because these steps do not need to be executed sequentially; they randomly sample a new location, exchange it with a current location in the subsample, and then evaluate the $\phi-$optimality criterion of the new subsample. So, if one wants to use larger values of $n_{cand}$, we recommend doing so in parallel. Increasing $n_{repeat}$ allows REX-SUB more chances to improve upon the subsample found in the previous search. However, in simulations and applications, it was found that the most substantial improvements in the subsamples (in terms of lowering the $\phi-$optimality criterion) occurred in the first few repetitions, with diminishing marginal returns. Therefore, if computational resources are limited, we recommend reducing $n_{repeat}$.

For the Vecchia approximation, we employ the maxmin ordering of locations, which has been shown to yield more accurate likelihood approximations and strong predictive performance, especially for cases with large number of locations \citep{guinness2018permutation}. Using maxmin ordering, we set the $m$ nearest neighbors of point $Z_{(i)} := Z(\bs_{(i)})$ based on the euclidean distance between locations. Selecting the size of the conditioning set (nearest neighbors) $m\ll n$ involves a tradeoff between the accuracy of the Vecchia approximation and the computational walltimes. We conduct a sensitivity analysis on the choice of $m$ in Section~\ref{sec:simuresults} to provide some practical guidance for specifying this tuning parameter. The standard Vecchia approximation is a straightforward approach where the joint density (\ref{eq:Vecchia}) is conditioned on the observed responses. 

 Sparse Gaussian process approximations such as the NNGP \citep{datta2016hierarchical} and Vecchia approximations \citep{vecchia1988estimation, katzfuss2020vecchia} introduce sparsity through restricted conditioning sets. The joint density is factorized into a product of conditional distributions, where the $i$-th observation is conditioned only on a subset of size $m$ (often its $m$ nearest neighbors) rather than the entire collection of previous observations. This construction results in a sparse precision matrix, which permits fast evaluation of the multivariate normal density and scalable inference for large spatial datasets. MRA \citep{katzfuss2017multi} represents the latent spatial process as a linear combination of basis functions specified at multiple spatial resolutions, where fine resolution captures local variability and coarser scales capture global behavior. MRA leverages recursive partitioning and the hierarchical basis representation to generate multi-resolution (block) sparse precision matrices, which enables scalable inference and distributed operations.
    
On the other hand, REX-SUB does not approximate the covariance or precision structure of a Gaussian process. Instead, it serves as a design-based subsampling strategy that selects curated subsets of locations using prediction-based objective functions. Scalability is attained by reducing the data size prior to model fitting, rather than by modifying or introducing sparsity into the underlying covariance structure.

\section{Simulation Study}
\label{sec:simuresults}
In this section, the algorithm introduced in Section \ref{sec:methods} is compared to random subsampling, LHS-based subsampling, and sequential IMSPE subsampling \citep{binois2021hetgp}. The sequential IMSPE samples are found using the \texttt{hetGP} package in R, using an LHS subsample with 9 points as the initial subsample. These subsamples are compared based on their mean squared prediction error, interval scores, and prediction interval coverage on simulated datasets.

\subsection{Study Design and Implementation}
\label{subsec:simulation_design}
This study aims to evaluate the robustness and scalability of REX-SUB under various spatial dependence structures commonly encountered in geostatistical applications. The subsamples generated by REX-SUB are compared to those obtained via random sampling and Latin Hypercube Sampling (LHS). Specifically, we assess REX-SUB and its competing approaches on its predictive accuracy and interval scores. We examine performance across a variety of spatial dependence regimes by systematically varying key parameters of the Mat\'ern covariance function—smoothness, range, and signal-to-noise ratio. These simulation settings are designed to capture a broad range of realistic spatial dependence structures, from smooth to rough surfaces and from weak to strong spatial signals. Additionally, we investigate how subsample size impacts model performance across these regimes.

We generate $N=12,500$ locations on the unit square domain $\bs\in\mathcal{D}=[0,1]^2$ where $n_{train}=10,000$ are used to train the model and reserve the remaining $n_{validate}=2,500$ locations to validate our approaches. The same validation set of $n_{validate}$ locations was used to evaluate all subsamples found. To clarify, Algorithm \ref{Alg:optsubsample} only uses the data in the training set. For each subsample, before running Algorithm \ref{Alg:optsubsample}, $\mathcal{I}_{test}$ was formed by randomly selecting 1,000 locations from the $n_{train} = 10,000$ locations. Once $\mathcal{I}_{test}$ is selected, the same set $\mathcal{I}_{test}$ is used throughout each exchange step in Algorithm \ref{Alg:optsubsample}. This way, the subsample found by Algorithm \ref{Alg:optsubsample} only includes locations from the remaining 9,000 locations that are neither in $\mathcal{I}_{test}$, nor the validation set. For each subsample, the metrics of interest (MSPE, Interval Score, and coverage) were measured on the validation set for each subsampling method of interest. The coverage is the proportion of the 2500 prediction intervals that capture the corresponding responses in the validation set.

The data are generated from a Gaussian process with zero mean $\mu(\bs)=\bzero$ and covariance function $K(\bs,\bs^{\prime}; \bTheta)$ where $\bTheta=(\nu, \ssq, \phi, \tau^2)^{\prime}$. Using a full factorial design, we examine combinations of smoothness ($\nu\in \{0.5,1.5\}$), effective ranges (defined as the distance where covariance decays to 0.05) $\rho^* \in \{0.3, 0.6\}$, and noise proportions $(\tau^2/(\tau^2+\sigma^2)= \{0.01, 0.1\})$. For a given effective range $\rho^*$ and smoothness $\nu$, numeric search was used to find the value of $\phi$ that brought the Matern covariance function (evaluated at $||\textbf{s} - \textbf{s}'|| = \rho^*$) as close as possible to 0.05. We fix the marginal variance at $\ssq=1$ across all settings to avoid identifiability issues between $\ssq$ and $\phi$, since it has been shown that only the ratio $\ssq/\phi$ is identifiable, not the individual parameters \cite{Zhang01032004,ying1991asymptotic,stein1990uniform}. In addition, we study two different subsample sizes ($n \in \{25,36\}$). In total, we consider 16 distinct simulation scenarios, each defined by a unique combination of the smoothness, range, noise proportion, and subsample size parameters. For each scenario, we generate 100 independent replicates, resulting in 1,600 model fits per approach. There are 8 different combinations of $v, \rho^*,$ and $\frac{\tau^2}{\tau^2 + \sigma^2}$. These combinations are denoted as Settings 1 to 8, and are listed in Table \ref{tab:simsettings}.

\begin{table}[!ht]
\centering 
\small
\begin{tabular}{c|ccc}
\hline
Setting & $\nu$ & $\rho^*$ & $\frac{\tau^2}{\tau^2 + \sigma^2}$ \\ \hline
 1 & 0.5   & 0.3      & 0.01                              \\
2 & 1.5   & 0.3      & 0.01                               \\
3 & 0.5   & 0.6      & 0.01                                 \\
4 & 1.5   & 0.6      & 0.01                       \\
5 & 0.5   & 0.3      & 0.10                                \\
6 & 1.5   & 0.3      & 0.10                        \\
7 & 0.5   & 0.6      & 0.10                   \\
8 & 1.5   & 0.6      & 0.10                          \\\hline
\end{tabular}
\caption{Summary of Simulation Settings}
\label{tab:simsettings}
\end{table}

\subsection{Results}
\label{subsec:simulation_results}

First, a sensitivity analysis was performed to determine the effect of the number of nearest neighbors ($m$) on the MSPE, interval scores, and CPU times (in seconds). The tests were done for the case where $\nu = 0.5, \rho^* = 0.3$, $\frac{\tau^2}{\tau^2 + \sigma^2} = 0.01$, and the subsample size was $n = 25$. The number of nearest neighbors considered were $m = 5, 10, 15$. The results of the sensitivity analysis are shown in Table \ref{tab:sensitivity}.

\begin{table}[!ht]
\small
\begin{tabular}{l|lll|lll|lll}
\hline
   & \multicolumn{3}{c|}{MSPE}           & \multicolumn{3}{c|}{Interval Score} & \multicolumn{3}{c}{CPU (s)} \\ \hline
m  & Rand & LHS    & REX-SUB           & Rand     & LHS       & REX-SUB    & Rand   & LHS   & REX-SUB  \\ \hline
5  & 0.99   & 0.91   & 0.54              & 5.03       & 5.08      & 4.43       & 0.09   & 0.1               & 56.77   \\
   & (0.01) & (0.02) & (\textless{}0.01) & (0.08)     & (0.14)    & (0.04)     & (0.01) & (0.01)            & (0.30)  \\
10 & 0.97   & 0.91   & 0.51              & 4.96       & 5.05      & 4.54       & 0.13   & 0.15              & 75.67   \\
   & (0.01) & (0.01) & (\textless{}0.01) & (0.08)     & (0.12)    & (0.05)     & (0.01) & (\textless{}0.01) & (0.08)  \\
15 & 1.00   & 0.90   & 0.49              & 5.06       & 5.09      & 4.60       & 0.11   & 0.12              & 85.98   \\
   & (0.01) & (0.01) & (\textless{}0.01) & (0.08)     & (0.14)    & (0.04)     & (0.01) & (0.01)            & (0.57) \\ \hline 
\end{tabular}
\caption{Comparison of average MSPE, Interval Scores, and CPU time for $m = 5,10,15$ neighbors used in the Vecchia approximation for $n = 25, \nu = 0.5, \rho^* = 0.3$, $\frac{\tau^2}{\tau^2 + \sigma^2} = 0.01$. (Standard Errors of the mean are in parentheses.) REX-SUB used the $\phi_{MSPE}$ criterion to select subsamples.}
\label{tab:sensitivity}
\end{table}

The sensitivity analysis in Table \ref{tab:sensitivity} shows that neither the MSPE nor the interval scores significantly changed when the number of neighbors changed. Decreasing the number of neighbors $m$ to 5  increases the average MSPE, while increasing the number of neighbors to $m = 15$ slightly decreased the MSPE. However, increasing $m$ leads to increases in the CPU time. Therefore, $m = 10$ neighbors were chosen for the Vecchia approximation throughout the rest of the simulations. The average CPU time for the REX-SUB method (which includes finding the subsample and fitting the Vecchia GP) in this case is roughly 76 seconds. An additional sensitivity analysis was performed to study the effect of changing the size of $\mathcal{I}_{test}$ on the performance metrics of REX-SUB. These results are summarized in Table S1 of the Supplementary Materials.

\begin{table}[!ht]
\tiny
\centering 
\setlength{\tabcolsep}{2.5pt} 
\begin{tabular}{c|llll|llll|llll}
\hline
        & \multicolumn{4}{c|}{MSPE}                    & \multicolumn{4}{c|}{Interval Score} & \multicolumn{4}{c}{Coverage}                 \\ \hline
Setting & Random & LHS    & hetGP  & REX-SUB           & Random  & LHS    & hetGP  & REX-SUB & Random & LHS    & hetGP  & REX-SUB           \\ \hline
1                            & 1.00   & 0.90              & 0.95   & 0.51              & 5.08    & 5.02   & 4.78   & 4.59    & 0.89   & 0.89              & 0.92              & 0.98              \\
                             & (0.01) & (0.01)            & (0.02) & (\textless{}0.01) & (0.09)  & (0.11) & (0.07) & (0.04)  & (0.01) & (0.01)            & (0.01)            & (\textless{}0.01) \\
2                            & 0.70   & 0.58              & 0.63   & 0.30              & 4.15    & 4.12   & 3.89   & 3.69    & 0.91   & 0.89              & 0.96              & 0.98              \\
                             & (0.02) & (0.01)            & (0.01) & (\textless{}0.01) & (0.09)  & (0.10) & (0.03) & (0.04   & (0.01) & (0.01)            & (\textless{}0.01) & (\textless{}0.01) \\
3                            & 0.59   & 0.55              & 0.60   & 0.27              & 4.13    & 4.08   & 4.56   & 3.73    & 0.90   & 0.89              & 0.87              & 0.98              \\
                             & (0.01) & (0.01)            & (0.01) & (\textless{}0.01) & (0.08)  & (0.09) & (0.13) & (0.04)  & (0.01) & (0.01)            & (0.01)            & (\textless{}0.01) \\
4                            & 0.28   & 0.21              & 0.31   & 0.10              & 2.38    & 2.17   & 2.73   & 2.07    & 0.92   & 0.91              & 0.97              & 0.98              \\
                             & (0.01) & (\textless{}0.01) & (0.01) & (\textless{}0.01) & (0.05)  & (0.04) & (0.03) & (0.02)  & (0.01) & (0.01)            & (\textless{}0.01) & (\textless{}0.01) \\
5                            & 0.99   & 0.99              & 0.97   & 0.59              & 5.08    & 4.93   & 5.23   & 4.88    & 0.90   & 0.91              & 0.89              & 0.98              \\
                             & (0.01) & (0.01)            & (0.01) & (\textless{}0.01) & (0.07)  & (0.05) & (0.14) & (0.04)  & (0.01) & (\textless{}0.01) & (0.01)            & (\textless{}0.01) \\
6                            & 0.57   & 0.54              & 0.60   & 0.31              & 3.88    & 3.93   & 4.33   & 3.77    & 0.91   & 0.89              & 0.88              & 0.98              \\
                             & (0.01) & (0.01)            & (0.01) & (\textless{}0.01) & (0.06)  & (0.07) & (0.13) & (0.04)  & (0.01) & (0.01)            & (0.01)            & (\textless{}0.01) \\
7                            & 0.59   & 0.56              & 0.59   & 0.34              & 4.01    & 4.08   & 4.24   & 3.56    & 0.90   & 0.87              & 0.88              & 0.98              \\
                             & (0.01) & (0.01)            & (0.01) & (\textless{}0.01) & (0.09)  & (0.10) & (0.12) & (0.03)  & (0.01) & (0.01)            & (0.01)            & (\textless{}0.01) \\
8                            & 0.45   & 0.39              & 0.43   & 0.21              & 3.43    & 3.26   & 3.18   & 3.07    & 0.90   & 0.90              & 0.94              & 0.97              \\
                             & (0.01) & (0.01)            & (0.01) & (\textless{}0.01) & (0.07)  & (0.06) & (0.02) & (0.01)  & (0.01) & (0.01)            & (0.01)            & (\textless{}0.01) \\ \hline
\end{tabular}
\caption{Comparison of Average MSPE, Interval Scores, and coverage over 100 replicated subsamples of size $n = 25$ from a sample of size $N = 10,000$. (Standard Errors of the mean are in parentheses.) REX-SUB used the $\phi_{MSPE}$ criterion to select subsamples.}
\label{tab:simresn25}
\end{table}

Tables \ref{tab:simresn25} and \ref{tab:simresn36} show the average MSPEs, Interval Scores, and coverage of prediction intervals on the validation set over 100 repeated subsamples of size $n = 25$ and $n = 36$, respectively, for the random, LHS, hetGP (with the IMSPE criterion), and REX-SUB (with the $\phi_{MSPE}$ criterion) methods. Tables \ref{tab:simresn25} and \ref{tab:simresn36} clearly show that in all 16 scenarios, the average MSPE and interval scores are lowest for the subsamples constructed by REX-SUB. Additionally, the prediction intervals for the REX-SUB method maintain at least 97\% coverage over the validation set across all 16 scenarios. REX-SUB consistently had the highest coverage among all four methods in 15 of the 16 examined scenarios. The exception was in Setting 4 when $n = 36$, where the hetGP method had 98\% coverage, and REX-SUB had 97\% coverage. In this case, REX-SUB had lower mean interval scores than hetGP, suggesting that it achieves similar coverage rates with narrower prediction intervals. Low MSPE values indicate that the GP fitted to the final subsample provides accurate predictions at test locations. Furthermore, lower average interval scores suggests that the GPs fit to the subsamples found by Algorithm \ref{Alg:optsubsample} systematically return narrower prediction intervals that capture the test location's values, compared to the other methods.

\begin{table}[!ht]
\tiny
\centering 
\setlength{\tabcolsep}{2.5pt} 
\begin{tabular}{c|llll|llll|llll}
\hline
        & \multicolumn{4}{c|}{MSPE}                    & \multicolumn{4}{c|}{Interval Score} & \multicolumn{4}{c}{Coverage}                 \\ \hline
Setting & Random & LHS    & hetGP  & REX-SUB           & Random  & LHS    & hetGP  & REX-SUB & Random & LHS    & hetGP  & REX-SUB           \\ \hline
1       & 0.84   & 0.78              & 0.82   & 0.46              & 4.6     & 4.65   & 4.3    & 4.29    & 0.91              & 0.89   & 0.93              & 0.99              \\
        & (0.01) & (0.01)            & (0.02) & (\textless{}0.01) & (0.08)  & (0.09) & (0.06) & (0.04)  & (0.01)            & (0.01) & (\textless{}0.01) & (\textless{}0.01) \\
2       & 0.56   & 0.45              & 0.48   & 0.24              & 3.71    & 3.46   & 3.51   & 3.43    & 0.91              & 0.92   & 0.97              & 0.99              \\
        & (0.01) & (0.01)            & (0.01) & (\textless{}0.01) & (0.06)  & (0.05) & (0.02) & (0.03)  & (0.01)            & (0.01) & (\textless{}0.01) & (\textless{}0.01) \\
3       & 0.52   & 0.47              & 0.53   & 0.24              & 3.69    & 3.64   & 4.25   & 3.35    & 0.91              & 0.91   & 0.89              & 0.98              \\
        & (0.01) & (0.01)            & (0.01) & (\textless{}0.01) & (0.05)  & (0.07) & (0.16) & (0.02)  & (\textless{}0.01) & (0.01) & (0.01)            & (\textless{}0.01) \\
4       & 0.20   & 0.14              & 0.25   & 0.08              & 1.95    & 1.77   & 2.50   & 1.79    & 0.92              & 0.92   & 0.98              & 0.97              \\
        & (0.01) & (\textless{}0.01) & 0.01   & (\textless{}0.01) & (0.03)  & (0.03) & (0.02) & (0.02)  & (0.01)            & (0.01) & (\textless{}0.01) & (\textless{}0.01) \\
5       & 0.91   & 0.88              & 0.87   & 0.53              & 4.69    & 4.76   & 4.60   & 4.58    & 0.91              & 0.90   & 0.92              & 0.98              \\
        & (0.01) & (0.01)            & (0.01) & (\textless{}0.01) & (0.05)  & (0.06) & (0.08) & (0.03)  & (\textless{}0.01) & (0.01) & (\textless{}0.01) & (\textless{}0.01) \\
6       & 0.52   & 0.49              & 0.55   & 0.27              & 3.57    & 3.66   & 4.10   & 3.52    & 0.92              & 0.90   & 0.89              & 0.98              \\
        & (0.01) & (0.01)            & (0.01) & (\textless{}0.01) & (0.03)  & (0.07) & (0.15) & (0.03)  & (\textless{}0.01) & (0.01) & (0.01)            & (\textless{}0.01) \\
7       & 0.52   & 0.49              & 0.53   & 0.31              & 3.56    & 3.60   & 3.91   & 3.42    & 0.91              & 0.90   & 0.89              & 0.99              \\
        & (0.01) & (0.01)            & (0.01) & (\textless{}0.01) & (0.05)  & (0.07) & (0.12) & (0.03)  & (\textless{}0.01) & (0.01) & (0.01)            & (\textless{}0.01) \\
8       & 0.38   & 0.30              & 0.34   & 0.18              & 2.93    & 2.94   & 2.82   & 2.78    & 0.92              & 0.90   & 0.96              & 0.98              \\
        & (0.01) & (\textless{}0.01) & (0.01) & (\textless{}0.01) & (0.04)  & (0.05) & (0.06) & (0.02)  & (\textless{}0.01) & (0.01) & (\textless{}0.01) & (\textless{}0.01) \\ \hline
\end{tabular}
\caption{Comparison of Average MSPE, Interval Scores, and coverage over 100 replicated subsamples of size $n = 36$ from a sample of size $N = 10,000$. (Standard Errors of the mean are in parentheses.) REX-SUB used the $\phi_{MSPE}$ criterion to select subsamples.}
\label{tab:simresn36}
\end{table}

When the subsample size increases from $n = 25$ to $n = 36$, the average MSPEs and interval scores decrease for all subsampling methods. This behavior is intuitive, because having more data points available generally leads to better model fits and narrower prediction intervals. When the error ratio $\tau^2/(\tau^2 + \sigma^2)$ increases from 0.01 to 0.10, the MSPE and interval score metrics tend to increase; this is especially most noticeable for the LHS and REX-SUB methods in Table \ref{tab:simresn36}. This behavior is also expected, since $\tau^2$ is the variance of the nugget in the covariance function (\ref{eq:matern}), and increasing this error variance increases the overall level of noise in the response. When the smoothness parameter $\nu$ increased from 0.5 to 1.5, the average MSPEs and interval scores decreased for all methods, and the average coverage tended to increase. As $\nu$ increases, the spatial random fields become progressively smoother and exhibits less abrupt local variation (i.e., less noise at shorter distances). Similar behavior appears for the average MSPEs and interval scores when the effective range $\rho^*$ increases from 0.3 to 0.6 in Tables \ref{tab:simresn25} and \ref{tab:simresn36}. When $\rho^*$ is smaller, the covariance function $K(\textbf{s},\textbf{s}'; \boldsymbol\Theta)$ decays faster for points $\textbf{s}$ and $\textbf{s}'$ that are further apart. With relatively low sample sizes of $n = 25, 36$, a smaller $\rho^*$ means that there are fewer informative points available to estimate the covariance function. Naturally, this implies that when $\rho^*$ increases, the model metrics improve for the models fit by all considered subsampling methods.

\begin{table}[!ht]
\centering 
\begin{tabular}{cclrrrr}
\hline
    &         &                                                  & \multicolumn{4}{c}{Average CPU (s)}                  \\ \hline
$n$ & Setting & \multicolumn{1}{l|}{Task}                        & Rand              & LHS  & hetGP & REX-SUB          \\ \hline
25  & 1       & \multicolumn{1}{l|}{Initial Sampling}            &  \textless{}0.01  & 0.02     & 0.01      &  \textless{}0.01                 \\
    &         & \multicolumn{1}{l|}{Fitting GP(s)}               &  0.11             &  0.11    & 0.25      &  35.83                 \\
    &         & \multicolumn{1}{l|}{Row Exchanges}             &                   &      &       &       0.02            \\
    &         & \multicolumn{1}{l|}{Finding Design Criterion} &                   &      &  21.47     &   45.38                \\
    &         & \multicolumn{1}{l|}{Total}                       &  0.11                 &  0.13    &   21.72    &   81.24                \\ \hline
25  & 2       & \multicolumn{1}{l|}{Initial Sampling}            & \textless{}0.01 & 0.02 & \textless{}0.01  & \textless{}0.01 \\
    &         & \multicolumn{1}{l|}{Fitting GP(s)}               & 0.09              & 0.08 & 0.17  & 30.81             \\
    &         & \multicolumn{1}{l|}{Row Exchanges}               &                   &      &       & 0.02              \\
    &         & \multicolumn{1}{l|}{Evaluating Criterion} &                   &      & 21.77 & 44.96             \\
    &         & \multicolumn{1}{l|}{Total}                       & 0.09              & 0.10 & 21.94 & 75.79             \\ \hline
36  & 1       & \multicolumn{1}{l|}{Initial Sampling}            &   \textless{}0.01 &  0.04  & 0.01  & \textless{}0.01   \\
    &         & \multicolumn{1}{l|}{Fitting GP(s)}               &   0.06            &  0.08  &  0.26 & 52.72              \\
    &         & \multicolumn{1}{l|}{Row Exchanges}               &                   &        &       & 0.03               \\
    &         & \multicolumn{1}{l|}{Evaluating Criterion} &                   &        & 37.82 & 65.51                  \\
    &         & \multicolumn{1}{l|}{Total}                       &   0.06            &  0.12  & 38.09 & 118.26            \\ \hline
36  & 2       & \multicolumn{1}{l|}{Initial Sampling}            &  \textless{}0.01  & 0.03 &  0.01 &  \textless{}0.01             \\
    &         & \multicolumn{1}{l|}{Fitting GP(s)}               &  0.07             & 0.08 &  0.25 &  53.82             \\
    &         & \multicolumn{1}{l|}{Row Exchanges}               &                   &      &       &  0.03                 \\
    &         & \multicolumn{1}{l|}{Evaluating Criterion} &                   &      & 37.36 & 64.47                  \\
    &         & \multicolumn{1}{l|}{Total}                       &  0.07             & 0.11 & 37.62 & 118.32             \\ \hline 
\end{tabular}
\caption{Average CPU time breakdown over 100 subsamples of size $n$ for each subsampling method in selected cases. All subsamples are taken from samples of size $N = 10,000$. Blank cells indicate that the subsampling method does not perform the corresponding task.}
\label{tab:CPUtimebreakdown}
\end{table}

Table \ref{tab:CPUtimebreakdown} shows a detailed breakdown of the average CPU time (in seconds) for each of the four considered subsampling methods. The runtime is broken down into four tasks: (i) initial sampling; (ii) fitting GPs; (iii) row exchanges; and (iv) evaluation of subsample criterion. Table \ref{tab:CPUtimebreakdown} also shows the total time, which is the sum of the time spent on these four tasks. Initial sampling is the cost of finding the first subsubsample via random sampling or random LHS subsampling (e.g. Step 1 of Algorithm \ref{Alg:optsubsample}). The time spent fitting GPs includes both fitting the GP model to the final subsample and the time spent fitting or updating GPs while subsampling. For REX-SUB, this includes Step 2 and the model-fitting portion of Step 5 in Algorithm \ref{Alg:optsubsample}. For hetGP, this includes updating the GP with the point that is found to minimize the IMSPE over all possible points in the sample. Only REX-SUB performs row exchanges, which corresponds to Steps 3, 4, 6, and 7 of Algorithm \ref{Alg:optsubsample}. These updates are generally very fast, since they only involve random sampling and updating row indices. The final task is evaluating subsample criterion. For hetGP, this includes the time spent evaluating the IMSPE criterion, and for REX-SUB, this includes the time spent computing and store $\phi^{(t)}$ in Step 5 of Algorithm \ref{Alg:optsubsample}. In all cases, REX-SUB had the longest CPU time, and it spent most of its time in Step 5, fitting GPs and evaluating the criterion $\phi_{MSPE}(\mathcal{I}^{(t)} \mid \textbf{S},\textbf{Z})$. This makes sense, because Algorithm \ref{Alg:optsubsample} requires fitting $n_{repeat} \times n \times n_{cand}$ GPs and evaluating the $\phi_{MSPE}$ criterion for each of these GPs on $\mathcal{I}_{test}$. In hetGP, the cost of updating GPs is mitigated due to fast closed-form $O(n)$ updates when replicates are added \citep{binois2021hetgp}, which is likely why the cost of fitting and updating GPs is so low in Table \ref{tab:CPUtimebreakdown}. Overall, these results suggest that there is a trade-off between computational speed and prediction accuracy. For example, although the hetGP method is faster than REX-SUB, the subsamples found by hetGP have higher MSPEs, higher interval scores, and lower coverage rates than those found by REX-SUB.

REX-SUB provides higher-than nominal coverage, signifying wider prediction intervals. On the other hand, random subsampling and LHS provide lower-than nominal coverage, suggesting narrower prediction intervals or biased predictions. The hetGP approach provides both under- and over-coverage depending on the setting. While REX-SUB exhibits slightly conservative coverage in some settings, the interval score \citep{gneiting2007strictly} provides a more comprehensive assessment by jointly evaluating coverage and interval width.

\section{Application}
\label{sec:example}

The National Aeronautics and Space Administration (NASA) launched the Terra satellite in December 1999 as part of the Earth Observing System. The Terra satellite orbits the Earth synchronous with the sun while collecting measurements of the Earth's atmosphere, land, and water. Terra carries a Moderate Resolution Imaging Spectroradiometer (MODIS), which is a remote-sensing monitor that captures measurements at high spatio-temporal resolutions. The Total Precipitable Water Daily L3 Global 1 Degree CMG dataset was acquired from the Level-1 and Atmosphere Archive \& Distribution System Distributed Active Archive Center, located in the Goddard Space Flight Center in Greenbelt, Maryland (\url{https://ladsweb.nascom.nasa.gov/}).

\begin{figure}[ht]
   \centering
\includegraphics[scale=.3]{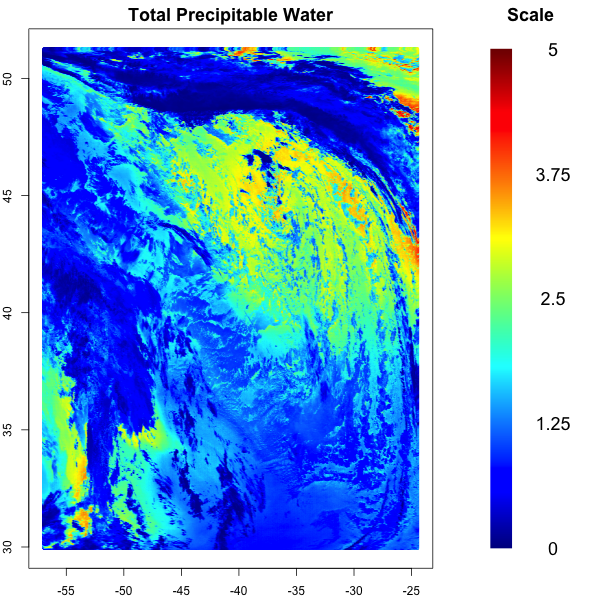} \\
   \caption{Total precipitable water data obtained via the MODIS Terra Satellite obtained on June 22, 2023. The x- and y-axes represent the longitude (-57.0 degrees to -24.5 degrees) and latitude (29.9 degrees to 51.3 degrees), respectively. Data were collected at 1 km$\times$ 1 km spatial resolutions over $2,748,620$ locations. The color scale represents total atmospheric column water vapor at each location.}
   \label{fig:precip}
\end{figure}

We use the MODIS level-2 atmospheric precipitable water product summarizing the total atmospheric column water vapor amounts over the earth. We focus on the region located at longitudes -57.0 degrees to -24.5 degrees, and latitudes 29.9 degrees to 51.3 degrees obtained on June 22, 2023 01:20:33UTC. The resolution of the data is 1 km$\times$ 1km with $N = 2,748,620$ locations. Values close to zero imply clear skies and larger values correspond to regions with more water vapor (see Figure \ref{fig:precip}). For this application, we model the log-transformed total precipitable water values to ensure the data have support on the real line.

\begin{figure}[!ht]
    \centering
    \includegraphics[width=0.5\linewidth]{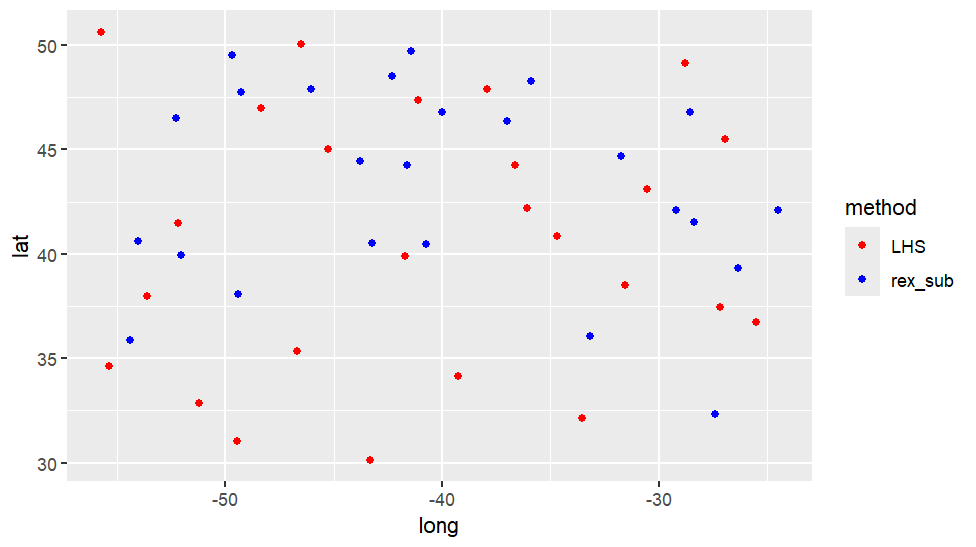}
    \caption{Example of Subsample Locations for REX-SUB and LHS when $n = 25$.}
    \label{fig:n25subsamplelocs}
\end{figure}

Random sampling, Latin Hypercube Sampling, sequential IMSPE sampling (using \texttt{hetGP}), and REX-SUB (with $n_{cand} = 10, n_{repeat} = 2$, and the $\phi_{MSPE}$ criterion) were used to find subsamples of size $n = 25$ and $n = 36$. The MODIS dataset was first randomly partitioned into a training set and a validation set; this was done by randomly selecting 10\% of the observations for validation. Each subsampling method was used to select subsamples of size $n$ from the training data. For each subsample, the MSPE and interval score metrics were computed on the validation data. This process was repeated 100 times for each method. An example of locations chosen by both REX-SUB and LHS is shown in Figure \ref{fig:n25subsamplelocs} for $n = 25$. 

\begin{figure}[!ht]
    \centering
    \includegraphics[width=0.75\linewidth]{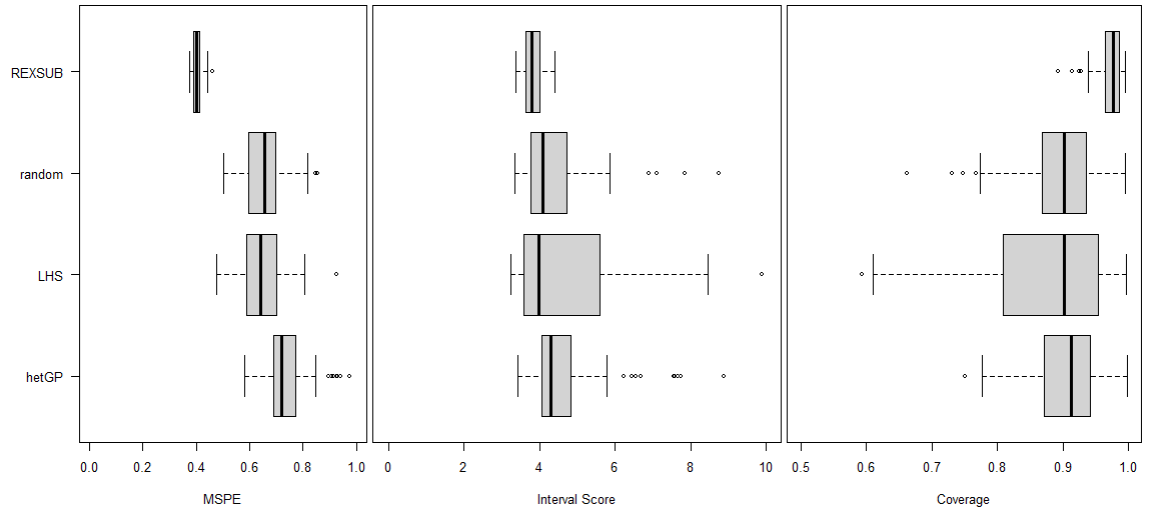}
    \caption{Comparison of MSPE, Interval Scores, and coverage of 100 subsamples of size $n = 25$ taken from the MODIS Data (full sample size $N = 2,748,620$).}
    \label{fig:boxplotsn25}
\end{figure}

\begin{figure}[!ht]
    \centering
    \includegraphics[width=0.75\linewidth]{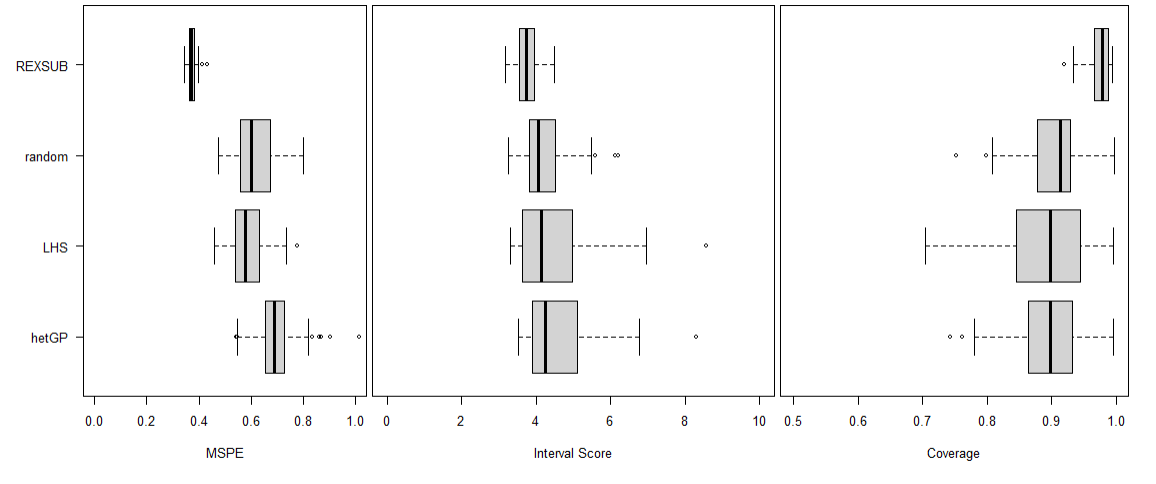}
    \caption{Comparison of the MSPE, Interval Scores, and coverage of 100 subsamples of size $n = 36$ taken from the MODIS Data (full sample size $N = 2,748,620$).}
    \label{fig:boxplotsn36}
\end{figure}

Figure \ref{fig:boxplotsn25} shows that for subsamples of size $n = 25$, the proposed REX-SUB method (with the MSPE criterion) had lower median MSPE and median interval scores on the validation dataset. The IQR of MSPEs and interval scores for the REX-SUB method are much lower than the other methods, suggesting that the REX-SUB method has more consistency in producing GPs with lower MSPEs and interval scores for the MODIS data. In terms of coverage, the REX-SUB method had higher median coverage than all other considered approaches. The IQR of the LHS method was the highest, suggesting that the coverage of prediction intervals provided by LHS fluctuates more than other methods. The coverage IQR for REX-SUB was lowest for $n = 25$ and $n = 36$, which suggests that REX-SUB consistently produces prediction intervals with high coverage on the MODIS dataset. Finding a single optimal subsample of size $n = 25$ takes an average of 6.9 minutes (414.2 seconds) using REX-SUB, and 0.11 minutes (6.5 seconds) using the \texttt{hetGP} package. Trying to fit a GP using the Vecchia approximation with $m = 10$ neighbors to the full dataset (with $N = 2,748,620$ observations) takes longer than 1 hour.

\section{Conclusion and Further Research Interests}
\label{sec:conc}

Modern spatial datasets have grown substantially in size, rendering standard Gaussian process models computationally prohibitive in practice. One promising solution to this problem is optimal subsampling, where an optimality criterion is used to select a smaller subset of the massive dataset to fit a GP. Existing subsampling methods for GPs typically focus on partitioning the space, assuming blockwise independence or filling the space as uniformly as possible. This study introduced REX-SUB, a randomized exchange algorithm that can identify subsamples with low mean squared prediction error and interval scores. REX-SUB is intended as a practical subsampling heuristic rather than a novel Gaussian process approximation. The objective is computational scalability while maintaining competitive inferential and predictive accuracy. The proposed optimality criteria (MSPE and interval scores) are evaluated by quickly fitting a GP to a small testing dataset. Starting with a random subsample, REX-SUB iteratively examines each location in the subsample and considers exchanging it with several random locations in the full dataset; REX-SUB then makes exchanges that yield the greatest reduction in the MSPE or the interval score. In Section \ref{sec:simuresults}, empirical results showed that subsamples selected using REX-SUB resulted in fitted GPs with lower MSPE and interval scores than GPs obtained from random and LHS subsamples. Compared to sequential IMSPE-based designs, which optimize predictive variance but may incur higher computational cost, REX-SUB provides a faster alternative that achieves a favorable accuracy–runtime tradeoff in large spatial datasets. In Section \ref{sec:example}, REX-SUB was applied to the MODIS dataset, which contains more than 2.7 million spatial locations. It was shown that REX-SUB is capable of finding small subsamples ($n = 25, 36$) that yielded lower prediction errors and interval scores than random and LHS subsampling.

There are two limitations to REX-SUB that warrant future research. Similar to the Federov exchange algorithm \citep{fedorov2013theory}, REX-SUB is not guaranteed to find a globally optimal subsample. This is because REX-SUB only makes exchanges that strictly improve the optimality criterion, and this can trap the algorithm in a local optimum. It would be interesting to consider other optimization algorithms that allow for sub-optimal exchanges to escape local optima, such as Simulated Annealing \citep{rios2025graphical, zheng2025exact} and Threshold Accepting \citep{rios2022ta}, which have both seen use in finding optimal experimental designs. Another limitation of REX-SUB is that a GP needs to be re-fit each time a location in the subsample is exchanged, which increases computation time. While this was mitigated by using the Vecchia approximation, it would be ideal to develop methods for quickly updating the optimality criterion without the need to re-fit the model. 

The current study is also subject to several limitations regarding spatial modeling. The spatial models assume a stationary and isotropic covariance function from the Mat\'ern class. Future extensions could explicitly model nonstationarity through weighted averaging of localized models \citep{fuentes2001high, kim2005analyzing, risser2015local}, basis function representations \citep{katzfuss2017multi, hefley2017basis}, or kernel convolution methods \citep{higdon1998process, paciorek2006spatial}. In addition, extending REX-SUB to the spatio-temporal setting would be useful. This could involve simple separable structures or more flexible approaches that account for interactions between space and time, including spatio-temporal basis functions \citep{zammit2021frk, cressie2008frk, cressie2011statistics}, multivariate bases, generalized additive models \citep{wood2004stable}, or nonseparable covariance models \citep{gneiting2002nonseparable, cressie1999classes}. Practical subsampling heuristics such as REX-SUB can further enhance scalable GP approximations by providing curated subsamples that the scalable approaches can use to fit their models. Hence, REX-SUB and the scalable GP methods may be more appropriately viewed as complementary components rather than direct competitors.

\section*{Data availability statement}
The authors confirm that the data supporting the findings of this study are available within the article and its supplementary materials.

\section*{Funding}

This research was not supported by any grants or funding. 

\section*{Disclosure statement}
No potential conflict of interest was reported by the authors.

\bibliographystyle{elsarticle-harv} 
\bibliography{biblo}

\newpage



%\begin{thebibliography}{00}


%\bibitem[Lamport(1994)]{lamport94}
%  \textit{\LaTeX: a document preparation system},
%  Addison Wesley, Massachusetts,

%\end{thebibliography}
\end{document}

\def\spacingset#1{\renewcommand{\baselinestretch}%
{#1}\small\normalsize} \spacingset{1}

\newtheorem{proposition}{Proposition}

\title{\bf Supplementary Material for \\ ``REX-SUB: A Scalable Subsampling Strategy for Modeling Large Spatial Datasets''}
\date{}
\maketitle

\section*{Sensitivity of REX-SUB to Size of $\mathcal{I}_{test}$}

An additional sensitivity analysis was performed to study the effect of changing the size of the test set $\mathcal{I}_{test}$ on the MSPE, interval scores, coverage, and CPU times of the proposed REX-SUB method. The tests were done for Setting 1, which is the case where $\nu = 0.5, \rho^* = 0.3$, $\frac{\tau^2}{\tau^2 + \sigma^2} = 0.01$, and the subsample size was $n = 25$. The test set $\mathcal{I}_{test}$ is found by randomly selecting a proportion $p \in (0,1)$ of the $N = 10,000$ training data points and setting them aside for evaluating the $\phi_{MSPE}$ criterion of the current subsample. None of the $pN$ points included in $\mathcal{I}_{test}$ are used for subsample optimization. 100 subsamples were found using the REX-SUB method with $n = 25$ and $|\mathcal{I}_{test}| = 10000p$ for $p = 0.05, 0.10, 0.15$.  
For each subsample, a GP was fit and the MSPE, Interval Scores, and average coverage were evaluated on the independent validation set of 2500 observations (generated separately from the $N = 10000$ observations).  The results of the sensitivity analysis are shown in Table \ref{tab:sensitivityItest}.

\renewcommand{\thetable}{S\arabic{table}}
\begin{table}[!ht]
\small
\centering 
\begin{tabular}{l|llll}
\hline
$p$    & MSPE                    & Interval Score                     & Coverage                  & CPU (s) \\ \hline
0.05 &  0.52                     & 4.59                               & 0.98                      &  68.15       \\
     &  (\textless{}0.01)        & (0.05)                             & (\textless{}0.01)         &  (3.58)       \\
0.10 &   0.51                    &  4.51                              &    0.98                   &  85.84        \\
     &   (\textless{}0.01)       &  (0.04)                            &  (\textless{}0.01)        &  (1.75)       \\
0.15 &   0.50                    &  4.62                              &  0.99                      &  96.68        \\
     &   (\textless{}0.01)       &  (0.04)                            & (\textless{}0.01)    &  (1.05)       \\ \hline
\end{tabular}
\caption{Comparison of average MSPE, Interval Scores, Coverage, and CPU time for test sets of size $|\mathcal{I}_{test}| = pN$ for $p = 0.05, 0.10, 0.15$, when $N = 10,000, n = 25, \nu = 0.5, \rho^* = 0.3$, $\frac{\tau^2}{\tau^2 + \sigma^2} = 0.01$. Metrics are averages over 100 subsamples. (Standard Errors of the mean are in parentheses.) REX-SUB used the $\phi_{MSPE}$ criterion to select subsamples.}
\label{tab:sensitivityItest}
\end{table}

Table \ref{tab:sensitivityItest} suggests that the MSPE and coverage of REX-SUB are not significantly impacted when the size of the test set changes; the MSPE decreases, but only slightly. The interval score slightly decreases when $p$ increases from 0.05 to 0.10, but then it increases when $p$ increases from 0.10 to 0.15. Increasing $p$ (which increases the size of $\mathcal{I}_{test}$ leads to increases in the average CPU time. This makes sense, because increasing the size of $\mathcal{I}_{test}$ directly increases the amount of time that it takes to evaluate $\phi_{MSPE}$, which is an average over all elements in $\mathcal{I}_{test}$. Based on these results, $p = 0.10$ was chosen as the proportion of observations for constructing $\mathcal{I}_{test}$, as it had moderate CPU times while achieving a lower mean interval score.

%The sensitivity analysis in Table \ref{tab:sensitivity} shows that neither the MSPE nor the interval scores significantly changed when the number of neighbors changed. Decreasing the number of neighbors $m$ to 5  increases the average MSPE, while increasing the number of neighbors to $m = 15$ slightly decreased the MSPE. However, increasing $m$ leads to increases in the CPU time. Therefore, $m = 10$ neighbors were chosen for the Vecchia approximation throughout the rest of the simulations. The average CPU time for the REX-SUB method (which includes finding the subsample and fitting the Vecchia GP) in this case is roughly 76 seconds. 

%\bibliographystyle{elsarticle-harv} 
%\bibliography{biblo}